\documentclass[journal, a4paper]{IEEEtran}

\usepackage{graphicx}   

\usepackage{psfrag}    
\usepackage{amssymb}
\usepackage{amsmath}   
\usepackage{caption}

\begin{document}

\title{Channel Capacity for MIMO Spectrally Precoded OFDM with Block Reflectors}
\author{Ghassen Zafzouf \\ The University of Queensland
\thanks{Advisor: Dr. Konstanty Bialkowski (UQ), Dr. Vaughan Clarkson (UQ).}}
\markboth{Draft 1}{}
\maketitle

\begin{abstract}
In contrast to the state-of-the-art technology, it is envisaged that 5G systems will accommodate a broad range of services and a variety of emerging applications. In fact, the advent of the Internet of Things (IoT) and its integration with conventional human-initiated transmissions yields the need for a fundamental system redesign. Hence, the ideal waveform for the upcoming 5G radio access technology should cope with several requirements, such as low computational complexity, good time and frequency localization and straightforward extension to \emph{multiple-input multiple-output} (MIMO) systems. In the light of reducing the algorithmic complexity of the spectrally precoded MIMO OFDM, we provide an in-depth analysis of the use of block reflectors. The channel capacity  formula for MIMO spectrally precoded OFDM is then derived, and the proposed design techniques are compared to conventional MIMO OFDM from an information-theoretic perspective.
\end{abstract}

\section{Introduction}
Unlike previous generations of mobile networks, 5G is expected to support new communication paradigms such as Device-to-Device (D2D) or Massive Machine-Type Communication (mMTC)  , and hence, it will lead to a fundamental shift of the state-of-the-art technology. The ongoing research activities have revealed the main drivers of the 5G technology, which are listed below. 
\begin{itemize}
  \item \textbf{Gigabit Wireless Connectivity -} 5G technology is envisioned to support the so-called immersive experience, which is enriched by \textit{context information}, \textit{ultra high defition video} and even \textit{augmented reality}. Therefore, 5G mobile networks will have to support very large data rates, i.e. access data rates in the range of Gbps.
  
  \item \textbf{Internet of Things (IoT) $\&$ Uncoordinated Access -} Motivated by the expected tremendous increase in the number of connected devices in the near future, researchers have been thoroughly investigating new communication concepts, particularly the so-called \textit{Internet of Things} (IoT) and \textit{Machine-to-Machine} (M2M) communication. This technology will lead to a ubiquitous connectivity of a large number of smart devices, which will enable some extremely advanced applications like smart grids, intelligent transportation systems and smart cities. The major challenge in this scenario is posed by the dramatic increase of the number of simultaneously connected objects compared to existing wireless systems.  
    
  \item \textbf{Tactile Internet -} A further requirement for 5G wireless cellular system is the ability to support a vast range of real-time applications characterized by extremely low-latency communications. The major motivation for this revolutionary change with respect to current LTE networks, is the tactile sense of the human body which can detect latencies in the order of $1$ ms accuracy. This feature of 5G will allow the emergence of mission-critical applications such as smart cities, telemedicine and public safety.
  
   \item \textbf{Cloud Computing -} As the concept of \emph{cloud services} enables the users to experience desktop-like applications requiring high computational cost in terms of power and storage while using mobile devices, the volume of mobile traffic will certainly increase and cellular networks will witness a massive exchange of information between the cloud and the devices. Therefore, all of the previously discussed technical requirements, i.e. very high data rate, minimal end-to-end delays and massive multiple access, are necessary for the emergence of this technology.     
\end{itemize}
Since any 5G communication system has to be able to exploit the benefits of multiple transmit and receive antennas, transmission diversity is a decisive feature for future wireless networks in order to achieve the required reliability and robustness under frequency-selective and time-variant channels. In Section \ref{MIMO Channel Modeling}, we start by presenting the employed MIMO channel models. Next, standard MIMO OFDM system is described in Section \ref{MIMO OFDM System Model} and the relevant parts of the corresponding transceiver are thoroughly analyzed. Then, we extend the investigated system by shaping the subcarriers in the frequency domain by means of an array of properly designed spectral precoders in Section \ref{MIMO Spectrally Precoded OFDM System Model}. Finally, the implementation details are presented in Section \ref{Numerical Results} and the system performance is assessed in both cases. Finally, Section \ref{Conclusions} concludes the paper.  

\section{MIMO Channel Modeling}
\label{MIMO Channel Modeling}
\begin{figure}[!hbt]
	\begin{center}
	\includegraphics[width=1\columnwidth]{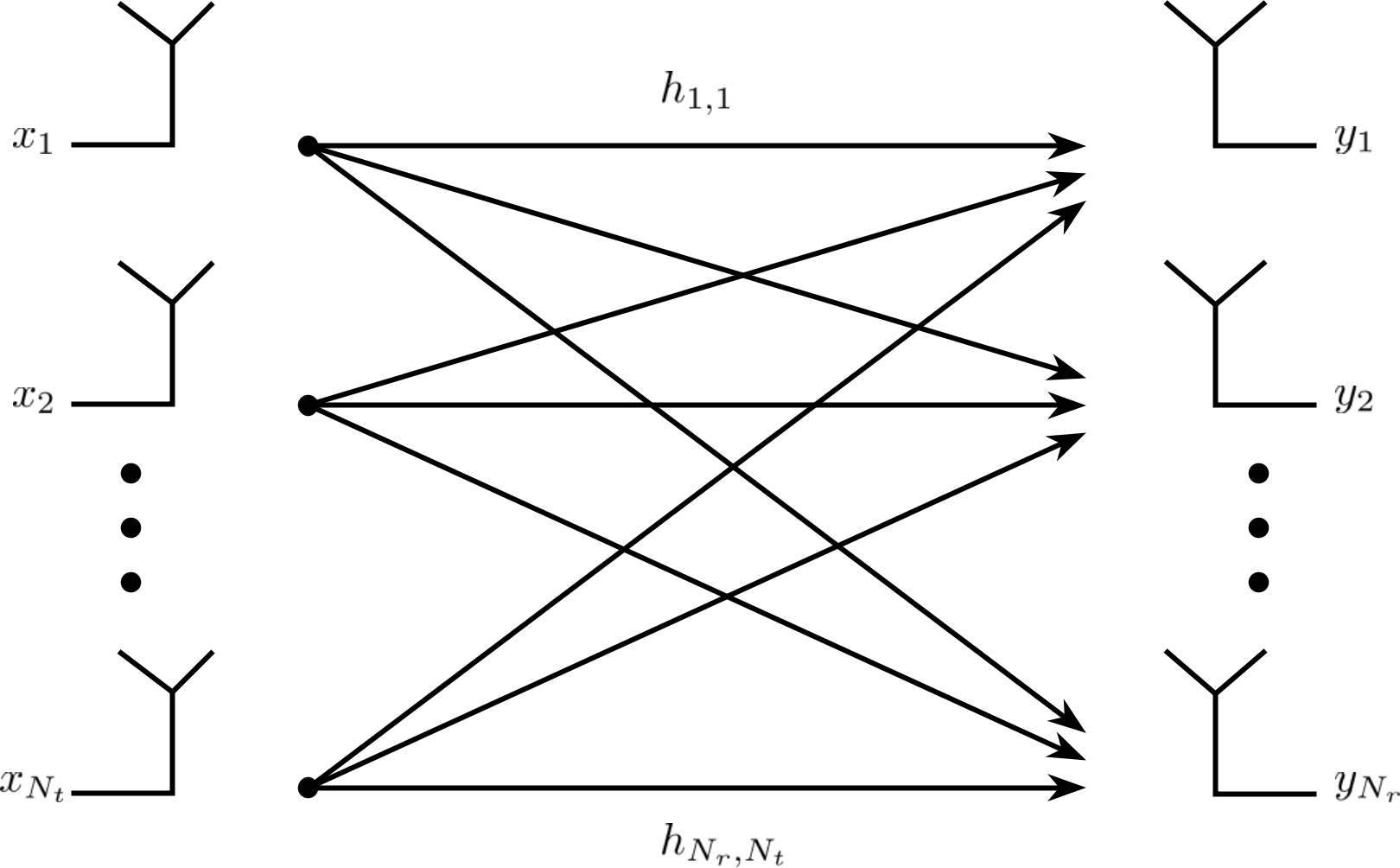}
	\end{center}
	\caption{Illustration of MIMO communication system.}
	\label{fig:MIMO_Channel_Model}
\end{figure}

The performance of a \emph{multiple-input multiple-output} (MIMO) is critically dependent on the availability of independent multiple channels. In fact, channel correlation will downgrade the performance of a MIMO system, especially its capacity. Channel correlation is a measure of similarity or likeliness between the
channels. In the extreme case that if the channels are fully correlated, then the MIMO system will have no difference from a single-antenna communication system. A narrowband (flat-fading) point-to-point communication system employing $N_t$ transmit and $N_r$ receive antennas is shown in Fig. \ref{fig:MIMO_Channel_Model}. \par
The channel capacity of a MIMO system does not only depend on the number of channels $(N_r \times N_t)$, but also depends on the correlation between the channels. In general, the greater the channel correlation is, the smaller is the channel capacity, as it will be shown in the next sections. The channel correlation of a MIMO system is
mainly due to the following two components: 
	\begin{enumerate}
		\item Spatial correlation
		\item Antenna mutual coupling
	\end{enumerate}
In a practical multipath wireless communication environment, the wireless channels are not independent from each other but due to scatterings in the propagation paths, the channels are related to each other with different degrees. This kind of correlation is called \emph{spatial correlation}. For a given channel matrix $\boldsymbol{H}$, the spatial correlation coefficients between the channels are defined as follows
	\begin{align}
\rho_{ij,pq} &= \frac{\mathrm{E}\left[h_{ij} h_{pq}^{*} \right]}{\sqrt{\mathrm{E}\left[h_{ij} h_{ij}^{*} \right] \mathrm{E}\left[h_{pq} h_{pq}^{*} \right]}}, 
	\end{align}
	with $i,p = 1,2, \cdots, N_t $ and $j,q = 1,2, \cdots, N_r $. 
The spatial correlation depends on the multipath signal environment. Multipath signals tend to leave the transmitter in a range of angular directions known as \emph{angles of departure} (AOD) rather than a single angular direction. This is the same for the multipath signals arriving at the receiver (called \emph{angles of arrival} (AOA)). Usually, the spatial correlation increases when AOD and AOA are reduced and vice versa.
\begin{figure}[!hbt]
	\begin{center}
	\includegraphics[width=\columnwidth]{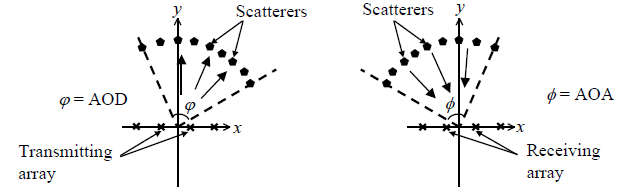}
	\end{center}
	\caption{(a) Angle of departure. (b) Angle of arrival.}
	\label{fig:AngleOfArrival_AngleOfDeparture}
\end{figure} 
\subsection*{Generation of a channel matrix $\boldsymbol{H}$ with specified spatial correlation}
If the channel correlation is known, we can use the method outlined in \cite{1011224} to generate the channel matrix $\boldsymbol{H}$ whose elements have the required correlation.
\begin{enumerate}
	\item Suppose $\boldsymbol{H}$ has the following form:
	\begin{align}
	 \boldsymbol{H} = \begin{bmatrix}
		h_{1,1} &h_{1,2} & \cdots &h_{1,N_t} \\
		h_{2,1} &h_{2,2} & \cdots &h_{2,N_t} \\
		\vdots &\vdots & \ddots &\vdots \\
		h_{N_r,1} &h_{N_r,2} & \cdots &h_{N_r,N_t} \\
		\end{bmatrix}.
	\end{align}
    \item Form the following vector $\mathrm{vec}(\boldsymbol{H})  \in \mathbb{C}^{N_r \cdot N_t}$ by stacking the column vectors of $ \boldsymbol{H} $ one by one:
    \begin{align}
	 \mathrm{vec}(\boldsymbol{H}) = \begin{bmatrix}
	h_{1,1} &\cdots &h_{N_r,1} &\cdots &h_{1,N_t} &\cdots &h_{N_r,N_t} \\
		\end{bmatrix}^{T}.
	\end{align}
	\item Compute the covariance matrix $\boldsymbol{R}_{\boldsymbol{H}}$ of $\mathrm{vec}(\boldsymbol{H})$ as follows 
	\begin{align}
		\boldsymbol{R}_{\boldsymbol{H}} = \mathrm{E}\left[ \mathrm{vec}(\boldsymbol{H}) \mathrm{vec}(\boldsymbol{H})^{H} \right]. 
	\end{align}
	\item Find the eigenvalues and eigenvectors of $\boldsymbol{R}_{\boldsymbol{H}}$.
	\item Then the channel matrix $\boldsymbol{H}$ can be expressed as
	\begin{align}
	 \mathrm{vec}(\boldsymbol{H}) &= \boldsymbol{V} \boldsymbol{D}^{\frac{1}{2}} \boldsymbol{r},
	 \label{eq:ComputeChannelMatrix}
	\end{align}	
 where the vector $\boldsymbol{r}$ contains i.i.d. complex Guassian random numbers with a unit variance and a zero mean, $\boldsymbol{V}$ is the matrix whose column vectors are the eigenvectors of
$\boldsymbol{R}_{\boldsymbol{H}}$, and $\boldsymbol{D}$ is a diagonal matrix whose diagonal elements are the eigenvalues of $\boldsymbol{R}_{\boldsymbol{H}}$. 
	\item Hence, once the desired correlation is given (by specifying $\boldsymbol{R}_{\boldsymbol{H}}$), the channel matrix $\boldsymbol{H}$ can be obtained using Eq. \ref{eq:ComputeChannelMatrix}.  
\end{enumerate}
In this paper, we compute the channel spatial correlation matrix $\boldsymbol{R}_{\boldsymbol{H}}$ as the Kronecker product of the individual spatial correlation matrices at the base station $\boldsymbol{R}_{BS}$ and at the mobile station $\boldsymbol{R}_{MS}$. These two matrices are defined for a $2 \times 2$ MIMO system as follows 
\begin{align}
\boldsymbol{R}_{BS} = \begin{bmatrix}
1 & \alpha\nonumber  \\
\alpha^{*} & 1 \\
\end{bmatrix},
\end{align}
and 
\begin{align}
\boldsymbol{R}_{MS} = \begin{bmatrix}
1 & \beta \nonumber  \\
\beta^{*} & 1 \\
\end{bmatrix}.
\end{align}
Hence the overall channel spatial correlation is 
\begin{align}
\boldsymbol{R}_{H} &= \boldsymbol{R}_{BS} \otimes \boldsymbol{R}_{MS} \nonumber \\
&= \begin{bmatrix}
1 & \beta & \alpha & \alpha \beta \nonumber \\
\beta^{*} &  1 & \alpha \beta^{*} & \alpha \nonumber \\
\alpha^{*} & \alpha^{*} \beta &1 & \beta \nonumber \\
\alpha^{*}\beta^{*} & \alpha^{*}  &\beta^{*} & 1 \nonumber \\ 
\end{bmatrix}
\end{align}
The parameters are defined by 3GPP in \cite{3GPP} for each correlation level as shown in Table \ref{tab_correlationPara}. 
\begin{table}[h]
	\centering
	\begin{tabular}{c || c c} 
	\hline
	\textbf{Correlation Level}           & $\alpha$ & $\beta$\\
	\hline
	\textbf{Low}           & 0   & 0    \\
	\textbf{Medium}        & 0.3 & 0.9  \\
	\textbf{High}          & 0.9 & 0.9       \\
	\hline
	\end{tabular}  
\caption{Correlation Factors}
\label{tab_correlationPara}
\end{table} 
\section{System Model}
\label{System Model}
\subsection{MIMO OFDM System Model}
\label{MIMO OFDM System Model}
\begin{figure}
	\centering	
	\includegraphics[width=\columnwidth]{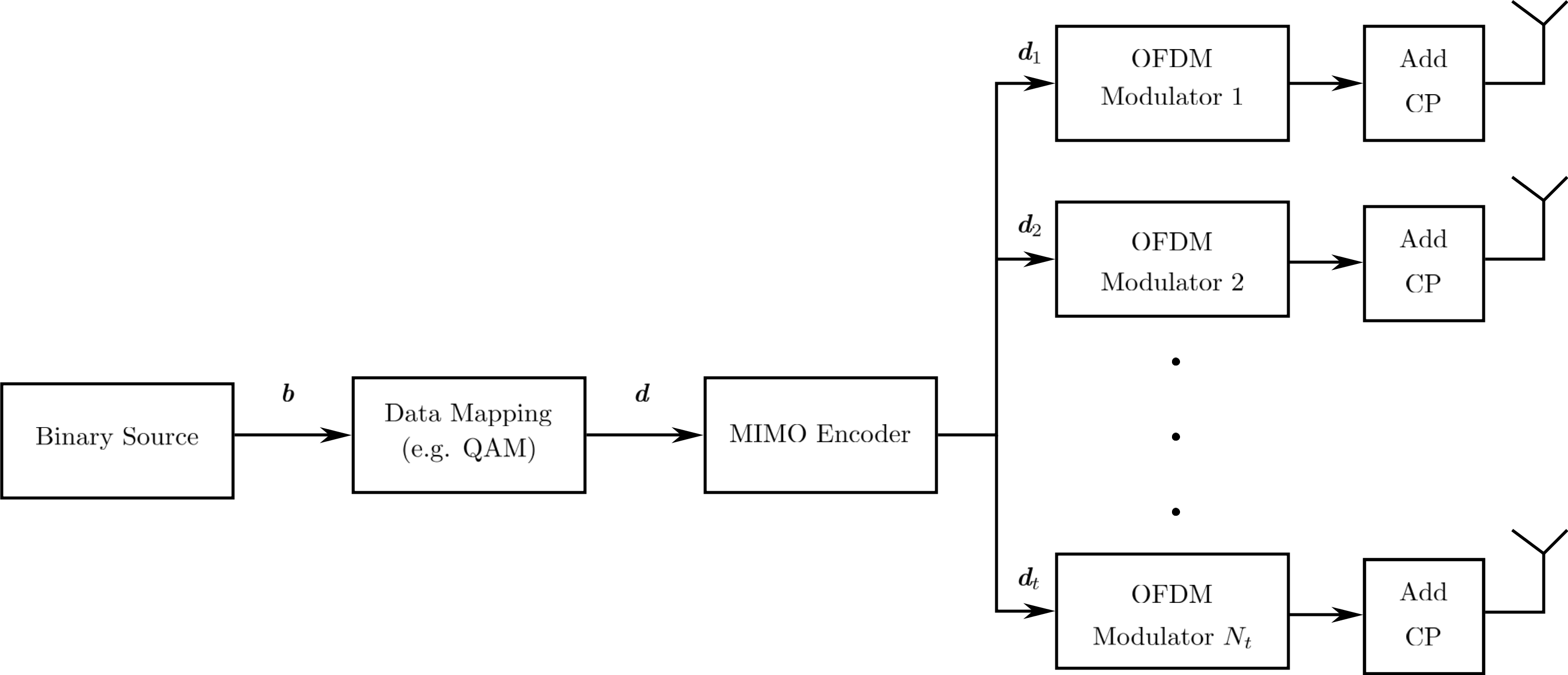}
	\caption{Schematic of a MIMO OFDM transmitter} 
	\label{fig:MIMO_OFDM_Tx}
 \end{figure}

\begin{figure}
	\centering	
	\includegraphics[width=\columnwidth]{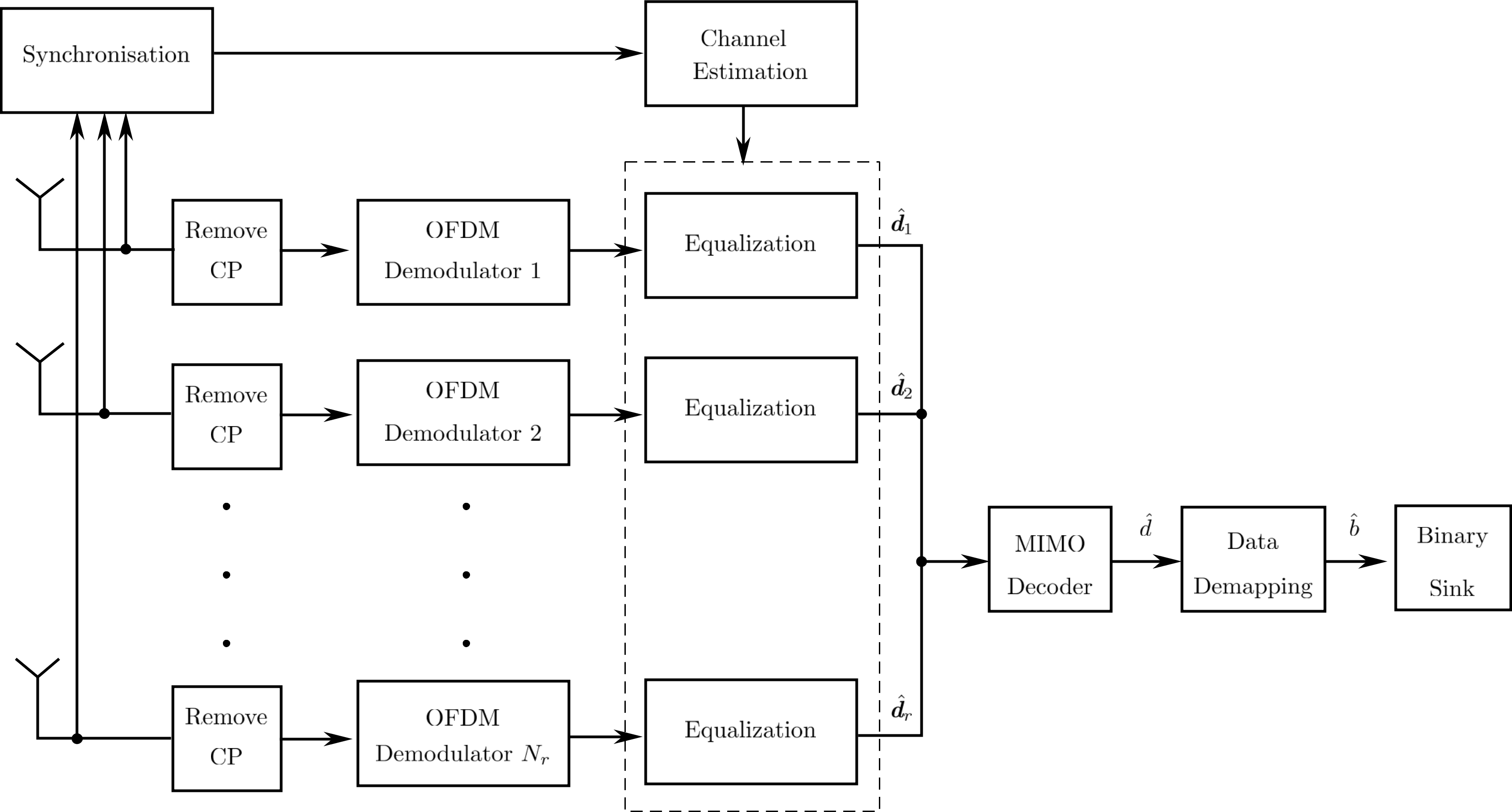}
	\caption{Schematic of a MIMO OFDM receiver} 
	\label{fig:MIMO_OFDM_Rx}
\end{figure}

\begin{figure}
	\centering	
	\includegraphics[width=\columnwidth]{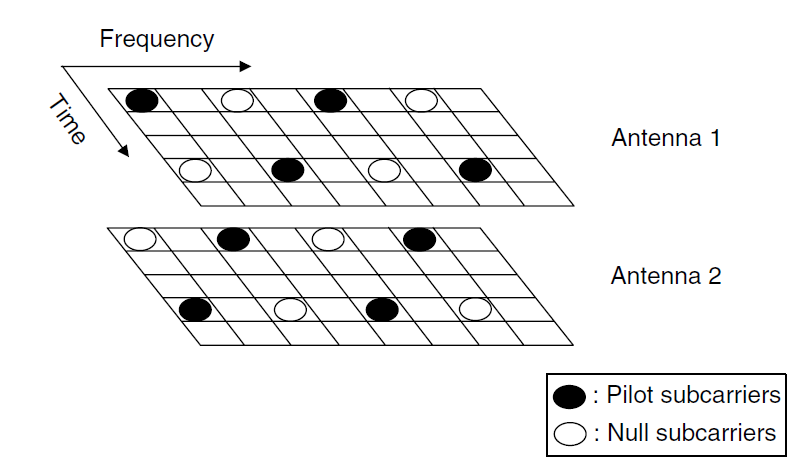}
	\caption{Pilot assignment for MIMO-OFDM scheme} 
	\label{fig:PilotAssignmentsMIMO}
 \end{figure}

We consider a MIMO OFDM communication system which employs $N_t$ transmit antennas, as shown in Fig. \ref{fig:MIMO_OFDM_Tx}, and $N_r$ receive antennas, as depicted in Fig. \ref{fig:MIMO_OFDM_Rx}. After mapping the binary message $\boldsymbol{b}$ into a stream of symbols $\boldsymbol{d}$ from a well-defined complex constellation such as QPSK or $M-$QAM, the MIMO encoder takes the modulated frequency-domain data vector and transforms it into $N_t$ parallel streams $\boldsymbol{d}_{i={1,2,\cdots,N_t}}$ to be further processed by the $i-$th OFDM modulator with $K$ subcarriers, as illustrated in Fig. \ref{fig:MIMO_OFDM_Tx}. For each subcarrier $ k=1, \cdots , K$ the $k$-th frequency-domain transmitted data vector $\boldsymbol{\Gamma}_k \in \mathbb{C}^{N_t}$ with covariance $\mathrm{E}\left[\boldsymbol{\Gamma}_k \boldsymbol{\Gamma}_l^{H}\right] = \dfrac{1}{N_t} \delta[k-l] \boldsymbol{I}_{N_t}$ can be expressed as following 
	\begin{align}
 		\boldsymbol{\Gamma}_k = \begin{bmatrix}
 		d_{1,k} &d_{2,k} &\cdots &d_{N_t,k}
		\end{bmatrix}   
	\end{align}
After the IFFT operation, the $n$-th OFDM symbol at the transmitter $i$ can be expressed as follows
	\begin{align}
		x_n^{i} &= \frac{1}{\sqrt{K}} \sum_{k=0}^{K-1}X_{k}^{i}e^{j\frac{2 \pi n k}{K}}
	\label{eq:OFDMSymbol}
	\end{align} 
Finally, in order to mitigate the effects of channel delay spread, a guard interval comprised of a cyclic prefix is appended to each sequence. The OFDM complex envelope is then obtained by passing the sequence through a DAC in order to generate the analog real and imaginary components, which are subsequently upconverted to an RF carrier frequency. The obtained analog signals are now transmitted simultaneously by the individual transmit antennas via a complex-valued random MIMO channel $\boldsymbol{H} \in  \mathbb{C}^{N_r \times N_t}$ with order $L_{ch}-1$ and characterized by the following transfer function
	\begin{align}
		\boldsymbol{H}\left(e^{j 2 \pi \theta}\right) = \sum_{l=0}^{L_{ch}-1}\boldsymbol{H}_le^{-j 2 \pi l \theta}, \qquad 0 \leq \theta \leq 1, 
	\end{align}
where the $N_r \times N_t-$dimensional complex-valued matrix $\boldsymbol{H}_l$ represents the $l$-th tap. Hence, the MIMO fading channel $\boldsymbol{H}_k$ on the $k-$th tone is defined as following 
	\begin{align}
		\boldsymbol{H}_k = \boldsymbol{H}\left(e^{j 2 \pi \dfrac{k}{K}}\right).
	\end{align}
Furthermore, we assume that the entries of $\boldsymbol{H}_k$ to be i.i.d. circularly symmetric complex Gaussian with zero mean and unit variance, i.e. $ \left[\boldsymbol{H}_k\right]_{i,j} \sim \mathcal{CN}(0,1)$ $\forall (i,j) $. Thus, the input-output relation of the downlink channel on the $k-$th subcarrier is given by
	\begin{align}
		\boldsymbol{y}_k &= \boldsymbol{H}_k \boldsymbol{\Gamma}_k + \boldsymbol{\eta}_k,
	\end{align} 
where $ \boldsymbol{y}_k = \begin{bmatrix}
y_1 &y_2 &\cdots &y_{N_r}
\end{bmatrix}$ is the $N_r$ received signal vector and $\boldsymbol{\eta}_k$ denotes the complex-valued circularly symmetric additive white Gaussian noise with the following covariance matrix 
\begin{align}
 \mathrm{E}\left[\boldsymbol{\eta}_k \boldsymbol{\eta}_l^{H}\right] =   \sigma_{\eta}^{2} \delta[k-l] \boldsymbol{I}_{N_r}. 
\end{align} 
 On the receiver side, before proceeding with OFDM demodulation and signal equalization, we firstly need to perform synchronisation. Therefore, a preamble comprising $N_t$ training symbols is appended to the transmitted signal. Usually the length of the guard interval in the training period is doubled, such as in IEEE802.16a, in order to allow frequency offset estimation and equalization in the special case where the length of the channel is larger than the guard interval length. Then, after discarding the CP and converting the received signal into frequency domain, the processed signal has to be equalized. Unlike \textit{single-input single-output} (SISO) OFDM systems the frequency-domain equalization in MIMO case does not only compensate the channel effects, like noise and ICI, but should also eliminate the inter-antenna interference and extract the superposed symbols. Thus, channel estimation for multiple antenna systems is not a straightforward extension of the single antenna systems. 
Firstly, we need to appropriately design and assign pilots for the MIMO OFDM system. Therefore, we use the grid-type pilot assignment procedure in both frequency and time domain as illustrated in Fig. \ref{fig:PilotAssignmentsMIMO}. 
This assignment scheme consists in sending a pilot symbol at a given subcarrier by one antenna while the other antennas remain silent, i.e. null subcarrier. The aim of this method is to avoid the inter-antenna interference, and thus, to protect each pilot from signals of the other antennas. Nevertheless, the downside of this scheme is the degradation of the system's spectral efficiency due to null subcarriers.  
\subsection{MIMO Spectrally Precoded OFDM System Model}
\label{MIMO Spectrally Precoded OFDM System Model}
\begin{figure}[!hbt]
	\begin{center}
	\includegraphics[width=\columnwidth]{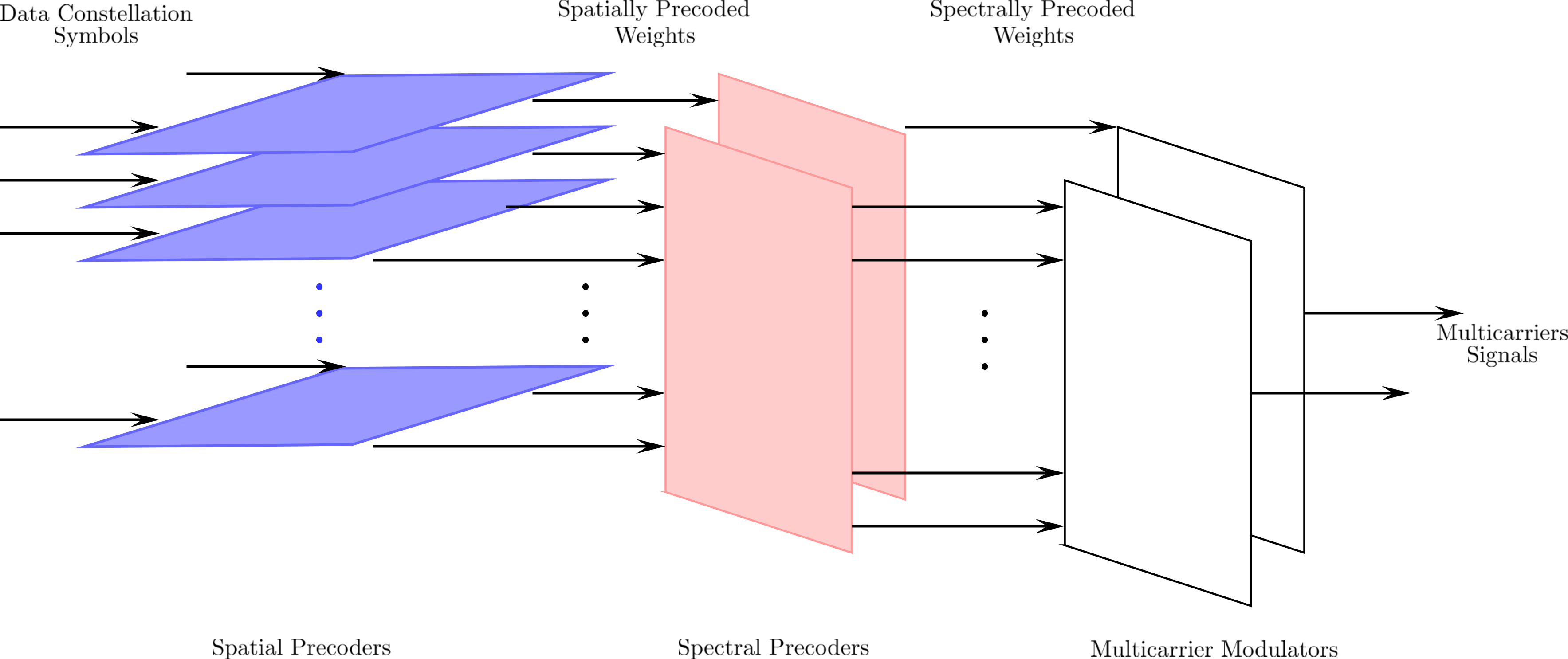}
	\end{center}
	\caption{Structure of spectrally precoded MIMO OFDM}
	\label{fig:Spectralyy_Precoded_MIMO_OFDM}
\end{figure} 
In order to reduce the OOB emission, the frequency-domain constellation symbols $\boldsymbol{d}_{j} =  \begin{bmatrix}
		d_{1,j} &d_{2,j} & \cdots &d_{L,j} \\
		\end{bmatrix}^{T} \in \mathbb{C}^{L}$ corresponding to the $j-$th transmit antenna are processed by a common spectral precoder $\boldsymbol{G}_{j}$ as follows
	\begin{align}
		\boldsymbol{\xi}_{j} = \boldsymbol{G}_{j} \boldsymbol{d}_{j},
		\label{eq:PrecodingOperation} 
	\end{align}		 
where the spectral precoder $\boldsymbol{G}_{j}$ is a $K \times L-$dimensional matrix generating the precoded frequency-domain data $\boldsymbol{\xi}_{j} =  \begin{bmatrix}
		\xi_{1,j} &\xi_{2,j} & \cdots &\xi_{K,j} \\
		\end{bmatrix}^{T} \in \mathbb{C}^{K}$. The structure of MIMO spectrally precoded OFDM is illustrated in Fig. \ref{fig:Spectralyy_Precoded_MIMO_OFDM}. Afterwards, the precoded symbols are then OFDM modulated and sent from antenna $j$. By organizing the data from frequency domain to the spatial domain, the corresponding MIMO transmission on the subcarrier $k$ at the receiver end can be expressed as follows
	\begin{align}
		\boldsymbol{y}_{k} &= 	\boldsymbol{H}_{k} \boldsymbol{\xi}_{k} + \boldsymbol{\eta}_{k}, 
		\label{eq:MIMOtransmissionMatrix}		
	\end{align}		   
with the vector $\boldsymbol{y}_{k}$ comprising the received symbols and $\boldsymbol{\eta}_{k}$ refers to the zero-mean additive white i.i.d Gaussian noise with covariance $ \mathrm{E}\left[ \boldsymbol{\eta}_{k} \boldsymbol{\eta}_{k}^{H} \right] = \sigma_{\boldsymbol{\eta}}^{2}\boldsymbol{I}_{N_r} $. Furthermore, the precoded symbols $\boldsymbol{\xi}_{k}$ from Eq. \ref{eq:PrecodingOperation} can be rewritten as a linear combination of the precoding matrix elements and the original data symbols over all subcarriers as following 
	\begin{align}
		\boldsymbol{\xi}_{k} = g_{k,k} \boldsymbol{d}_{k} + \sum_{j \neq k}^{L}	 g_{k,j} \boldsymbol{d}_{j}.	
	\end{align} 
Hence, the received data on the $k-$th tone can be expressed as 
	\begin{align}
		\boldsymbol{y}_{k} &=  g_{k,k} \boldsymbol{H}_{k} \boldsymbol{d}_{k} + \sum_{j \neq k}^{L} g_{k,j} \boldsymbol{H}_{k} \boldsymbol{d}_{j} + \boldsymbol{\eta}_{k}. 
		\label{eq:MIMOlinear}
	\end{align} 
By observing Eq. \ref{eq:MIMOlinear}, we can conclude that each coefficient of the spectral precoding matrix $\boldsymbol{G}_{k}$ has the effect to weight a $N_r \times N_t$ channel matrix. Thus, the overall MIMO transmission can be rewritten in one linear equation by introducing the following extended spectral precoder $\boldsymbol{\tilde{G}}$
	\begin{align}
		\boldsymbol{\tilde{G}} &= \boldsymbol{G}_{k} \otimes \boldsymbol{1}_{N_r,N_t},
	\end{align}
where the operator $\otimes$ refers to the Kronecker product and $\boldsymbol{1}_{N_r,N_t}$ is an $N_r \times N_t-$dimensional all-ones matrix. At this point, we define the data vector $\boldsymbol{y} = \left[\boldsymbol{y}_{1}^{T} \cdots \boldsymbol{y}_{K}^{T}  \right]^{T} \in \mathbb{C}^{N_rK} $ comprising all received symbols and the noise vector $\boldsymbol{\eta} = \left[\boldsymbol{\eta}_{1}^{T} \cdots \boldsymbol{\eta}_{K}^{T}  \right]^{T} \in \mathbb{C}^{N_rK}$ which contains the receiver Gaussian noise. Additionally, let $\boldsymbol{d} = \left[\boldsymbol{d}_{1}^{T} \cdots \boldsymbol{d}_{L}^{T}  \right]^{T} \in \mathbb{C}^{N_tL}$ denote the overall input signal. Hence, the input-output relation is equivalent to an overall MIMO transmission as follows
	\begin{align}
		\boldsymbol{y} = \boldsymbol{H}_{eff} \boldsymbol{d} + \boldsymbol{\eta},
		\label{eq:overallMIMOexpression}
	\end{align}
where the matrix $\boldsymbol{H}_{eff}$ corresponds to the overall effective channel defined as follows 
	\begin{align}
		\boldsymbol{H}_{eff} &= \boldsymbol{\tilde{G}} \odot \boldsymbol{H},
	\end{align}	      
with $\boldsymbol{H} = \left[ \boldsymbol{H}_{1}^{T} \cdots \boldsymbol{H}_{K}^{T} \right]^{T} \in  \mathbb{C}^{N_rK \times N_t} $ is the concatenation of per-subcarriers MIMO channels. Using the result from Eq. \ref{eq:overallMIMOexpression} we can study the capacity of the spectrally precoded OFDM system given different precoder design schemes. In fact, the ergodic capacity of such communication system can be expressed as follows 
  	\begin{align}
  		C &= \mathrm{E}\left[ \log_2 \left( \det \left(\boldsymbol{I}_{N_r} + \varrho \boldsymbol{H}_{eff}^{H}\boldsymbol{H}_{eff} \right) \right) \right],  
	\end{align}       

with $\varrho = \frac{1}{N_t \sigma_{\eta}^{2}}$ refers to the per-stream SNR.

\section{Numerical Results}
\label{Numerical Results}
\begin{figure}[!hbt]
	\begin{center}
	\includegraphics[width=1.1\columnwidth]{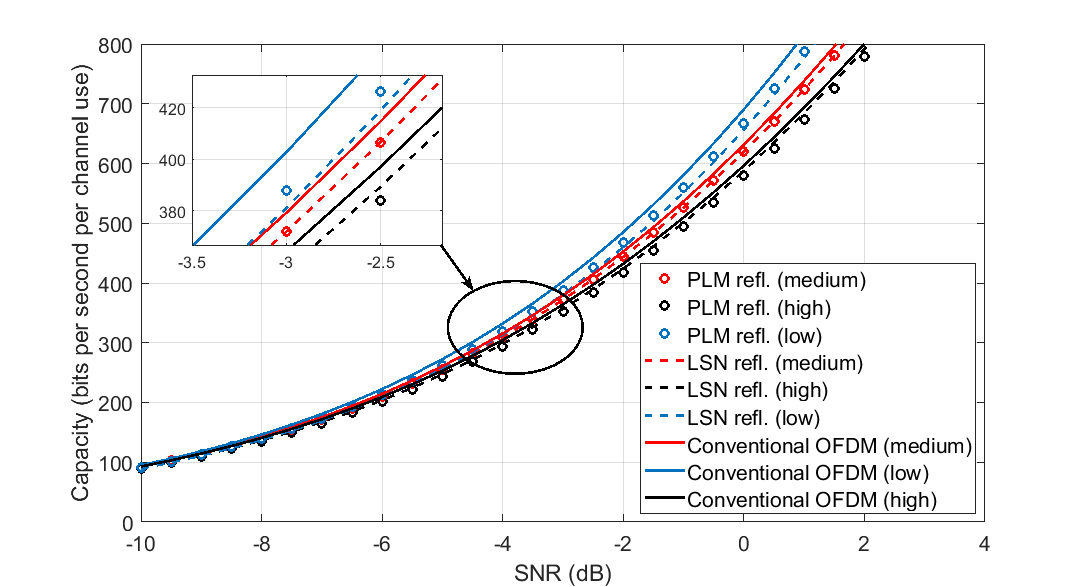}.
	\end{center}
	\caption{Channel capacity for $2 \times 2$ MIMO spectrally precoded and conventional OFDM systems}
	\label{fig:Capacity_comparison_Precoded}
\end{figure}

With the aim of gaining more insight into the impact of spectral precoder on MIMO OFDM communication system, we simulate three scenarios with the parameters as indicated in Table \ref{tab:SystemParameters}. These parameters are inspired by (4G) E-UTRA/LTE case \cite{LTE}. For means of comparison, we used two different precoding schemes: The first approach is known as \textit{Orthogonal Power Leakage Minimizing} (PLM) and was presented by Ma \textit{et al.} in \cite{5672368}. The second precoding technique is called \textit{Least-Squares Notching} (LSN) precoder, and was introduced by van de Beek in \cite{vdB}. We plot the channel capacity of spectrally precoded OFDM using LSN precoder and orthogonal PLM precoder with block reflectors compared to conventional OFDM in Fig. \ref{fig:Capacity_comparison_Precoded} for different correlation levels, i.e. low, medium and high. As it can be observed in this figure, employing a spectral precoder slightly reduces the channel capacity. Furthermore, at moderate SNR (starting from $0$ dB) the capacity lines of LSN and orthogonal PLM precoders are not exactly overlapping. Actually, in case of highly correlated channels the LSN precoder with block reflectors seems to achieve better performance than orthogonal PLM precoder. However, as the correlation decreases, the difference between the channel capacity associated with LSN-precoded OFDM and standard OFDM increases, whereas the capacity for PLM precoder is approximately the same as compared to conventional OFDM.
\begin{table*}[ht]
\centering
	\begin{tabular}{l c c c} 
	\hline
	\textbf{Parameter} & \textbf{Conventional OFDM} & \textbf{Spec. Pre. OFDM (LSN + refl.)} & \textbf{Spec. Pre. OFDM (PLM + refl.)}  \\ 
	\hline
	 Number of transmit antennas $N_t$ & 2 & 2  & 2      \\
	 Number of receive antennas $N_r$ & 2 & 2  & 2       \\
	 Available subcarriers & 2048 & 2048  & 2048      \\
	 Subcarrier spacing & $15$ kHz  & $15$ kHz  & $15$ kHz   \\ $\Delta f = \frac{1}{T_s} $  \\
	 Occupied subcarriers  & $\left[-300,300  \right] \setminus \lbrace 0 \rbrace $ & $\left[-300,300  \right] \setminus \lbrace 0 \rbrace $ & $\left[-300,300  \right] \setminus \lbrace 0 \rbrace $        \\
	 Cyclic prefix $(T_{cp})$ & $\frac{9}{128} T_s$ & $\frac{9}{128} T_s$  & $\frac{9}{128} T_s$        \\
	 Number of constraints  & None & 8 & 8 \\
	 Considered frequencies &$\emptyset$ &$\lbrace \pm 5100 \pm 1, \pm 6100 \pm 1 \rbrace$ kHz & $\left[ -40,-5  \right] \cup \left[ 5,40  \right]$ MHz \\ & & & sampled at $F_s = 200 $ kHz \\    
	\hline
	\end{tabular}
\caption{Communication system parameters}
\label{tab:SystemParameters}
\end{table*}
\section{Conclusions}
\label{Conclusions}
Since MIMO technology plays a central role in 5G technology and there has been no prominent effort in this direction, we have mainly focused on extending the spectrally precoded OFDM scheme with block reflectors to the MIMO case. In this context, the channel capacity formula for the adopted modulation technique was derived and the channel's correlation degree, i.e. low, medium or high, on the capacity was thoroughly discussed. For means of comparison, we investigated two different precoding approaches, namely orthogonal PLM and LSN precoders. Through a simulation inspired by the state-of-the-art technology (LTE), we have demonstrated that LSN precoder outperforms PLM over highly correlated channels, whereas PLM is the best strategy to adopt when the channel correlation degree is low.  

\begin{flushleft}

\end{flushleft}
\end{document}